\documentclass[%
 reprint,
 amsmath,amssymb,
 aps,
]{revtex4-1}

\usepackage{graphicx}% Include figure files
\usepackage{dcolumn}% Align table columns on decimal point
\usepackage{bm}% bold math
\usepackage{color}
\usepackage{xcolor}

\def \be {\begin{equation}}
\def \ee {\end{equation}}
\def \ba {\begin{array}}
\def \ea {\end{array}}
\def \p {\partial}

\begin{document}

\preprint{APS/123-QED}

\title{Semiclassical approach for excitonic spectrum of Coulomb coupling \\ between two Dirac particles}% Force line breaks with \\

\author{Victor Zalipaev}
\affiliation{ITMO University,  St. Petersburg, 197101, Russia}
\email{vvzalipaev@itmo.ru}%

% \altaffiliation[Also at ]{Physics Department, XYZ University.}%Lines break automatically or can be forced with \\
%\author{Vladislav Kuidin}%
\author{Vladislav Kuidin}%
\affiliation{ITMO University,  St. Petersburg, 197101, Russia}
%\\This line break forced with \textbackslash\textbackslash }%

\begin{abstract}
The properties of energy spectrum of excitons in monolayer
transition metal dichalcogenides are investigated using  a multiband
model. In the multiband model we use  the excitonic Hamiltonian in
the product base of the Dirac single-particle states at the
conduction and valence band edges.
Following the separation of variables we decouple the corresponding
energy eigenvalue system of the first order ODE radial equations
rigorously and solve the resulting the second order ODE
self-consistently, using the finite difference method, thus we
determine the energy eigenvalues of the discrete excitonic spectrum
and the corresponding wave functions. We also developed WKB approach
to solve the same spectral problem in semiclassical aproximation for
the resulting ODE. We compare the results for the energy spectrum
and the corresponding eigen-functions forms  for WS2 and WSe2
obtained by means of both methods. We also compare our results for
the energy spectrum  with other theoretical works for excitons, and
with available experimental data.

\end{abstract}

%\keywords{Suggested keywords}%Use showkeys class option if keyword
                              %display desired
\maketitle

%\tableofcontents

\section{\label{sec:level1} Introduction}

%\protect\\ The line break was forced \lowercase{via} \textbackslash\textbackslash}

%In this paper we present a theoretical analysis of the energy spectrum of excitonic states in the transition metal dichalcogenide (TMD)  monolayer.

Two-dimensional atomically thin 
graphene-type materials such as TMD monolayers ($MoS_2$, $MoSe_2$,
$WS_2$, $WSe_2$)  with the stochiometric formula $MX_2$, where
$M$ represents a transition metal, like Mo or $W$, and $X$ stands
for a chalcogenide ($S$, $Se$, or $Te$) attract a high interest due to the fact that they display new fundamental physical properties that are expected to be
important for future applications in electronics and optics
 with the emphasis in optoelectronics and photodetection, where
optical absorption plays a central role~\cite{Mak2010, Ramasubramaniam2012, Xiao2012, Kormnyos2015, Splendiani2010, Mak2012nano, Zeng2012, Cao2012, Sallen2012}. It is therefore of utmost
importance to understand the dominating optical absorption mechanism
in 2D TMDs, which has strong excitonic character . In contrast to
graphene, which has a gapless spectrum, the inversion
symmetry breaking in TMD monolayers leads to the formation of a
direct band gap. 
%Furthermore, only a few theoretical works describing the excitonic absorption spectrum of these materials were published recently (see for example the papers \cite{Berman1}, \cite{Berman2}, \cite{Trushin}, \cite{Peeters}). 
It is worth to mention that in \cite{Trushin} a
rigorous approach of separation of variables was developed for the
pure Coulomb potential that is opposite to the finite element
analysis presented in \cite{Peeters} which was applied to the case the screened Coulomb
potential - the Keldysh potential\cite{rytova1967the8248,keldysh,Cudazzo2011}.

Moreover, it was understood that
 for ultra thin semiconductors the dielectric
environment plays a crucial role and influences the effective
strength of the Coulomb potentials inside a semiconductor layer (see
\cite{Peeters}). Such long-range interactions become stronger as the
thickness of the semiconductor layer decreases, which allows the
formation of neutral and charged excitons. This
enhanced Coulomb interaction leads to high exciton binding energies of neutral~\cite{Mak2010, Splendiani2010, Komsa2012, Feng2012, Qiu2013, BenAmara2016} and charged excitons~\cite{Mak2012, Ross2014, Lui2014, Rezk2016, Zhang2014, Mouri2013, Singh2016, Scheuschner2014, Soklaski2014, Zhang2015_2} recently attracting a particular interest in the scientific community~\cite{Zhumagulov-trion-PRB,Zhumagulov-trion-PL,arora2019dark}.

%which, are $1-2$ orders of magnitude larger than excitons in typical semiconductors. 
%Numerous photoluminescence experiments in TMD monolayers  confirmed the existence of excitonic states that are localized in the band gap.

%An exciton is a bound state of an electron and a hole which are attracted to each other by the Coulomb force. The electron-hole pair in 2D semiconductors has usually been described as a 2D hydrogenlike system with the reduced mass. However, as it was discovered the exciton spectrum in 2D TMDs does not resemble the conventional Rydberg series.

In our paper the properties of energy spectrum of excitons in TMD
monolayer  are investigated using a multiband model. The starting
point of our analysis is the excitonic Hamiltonian constructed in
\cite{Peeters}. In this multiband model we use  the excitonic Hamiltonian
in the product base of the Dirac single-particle states at the
conduction and valence band edges described in the paper \cite{Peeters}.
This includes the effect of spin-orbit coupling, in the product base
of the single-particle states at the conduction and valence band
edges. Following the separation of variables we decouple the
corresponding energy eigenvalue system of the first order ODE radial
equations rigorously and solve the resulting the second order ODE
self-consistently using the finite difference method. Thus we
determine the energy eigenvalues of the descrete excitonic spectrum
and the corresponding wave functions. We also developed WKB approach
to solve the same spectral problem in the semiclassical
approximation for the resulting ODE in the case of the Keldysh
potential application. Similar to \cite{Trushin}, \cite{Peeters}, we compare the
results for the energy spectrum and the corresponding
eigen-functions forms for WS2 and WSe2 obtained by means of both
methods. We also compare our results for the energy spectrum  with the data presented in \cite{Trushin} for the excitonic s-states.

 According to
\cite{Peeters}, the exciton Hamiltonian is constructed in the basis
$(|\phi^e_c\rangle\otimes|\phi^h_c\rangle,
|\phi^e_c\rangle\otimes|\phi^h_v\rangle,|\phi^e_v\rangle\otimes|\phi^h_c\rangle,|\phi^e_v\rangle\otimes|\phi^h_v\rangle,)^T$,
that is the basis states spanning the total Hilbert space that is
given by the set of all the possible combinations of tensor products
of the atomic orbital states of the individual particles at the
conduction and valence band edges. Using the orthonormality of the
basis functions, the total exciton Hamiltonian (symbol of the
operator) could be written in the form introduced in \cite{Peeters} (see formulas (4) and (5))
with the Keldysh potential $-V(r)$ in polar coordinates that is given by~\cite{rytova1967the8248,keldysh,Cudazzo2011}
 \be
V(r) = V_0\frac{\pi}{2} \left(H_0\left(\frac{r}{r_0}\right)-Y_0\left(\frac{r}{r_0}\right)\right),
\label{Keldysh}
 \ee
where $Y_0(r)$ and $H_0(r)$ are the Newmann function and the Struve
function, respectively, $r_0$ is the screening length (the
characteristic space scale of the problem),
 \be
 V_0=\frac{e^2}{4\pi \epsilon_0 \chi r_0
 t},~~~\chi=\frac{\epsilon_1+\epsilon_2}{2},
 \ee
 where $t$ is the hopping parameter, $\epsilon_{1,2}$ is the dielectric constant
of the environment above  and below the TMD monolayer. It is worth
remarking that $ at=v_F\hbar$
where $v_F$ is the Fermi velocity and $a$ is the lattice constant.

The eigenvalue problem for this Hamiltonian
 \be
H^{exc}_{\alpha}(\pmb{k}^e,\pmb{k}^h,r_{eh})|\Psi_{\alpha,n}\rangle=E^{exc}_{\alpha,n}(\pmb{k}^e,\pmb{k}^h)|\Psi_{\alpha,n}\rangle
 \ee
defines the exciton energy
$E^{exc}_{\alpha,n}(\pmb{k}^e,\pmb{k}^h,r_{eh})$ and the exciton
eigenstate $|\Psi_{\alpha,n}\rangle$, where $\alpha$ is a notation for the set of the spin and the valley indexes $s_e,\tau_e$, $s_h,\tau_h$ with the values $\pm 1$ for the electron-hole Dirac particles, $\pmb{k}^e,\pmb{k}^h$  are their wave numbers, $n$ means a set of quantum numbers of the excitonic states of discrete spectrum (\cite{Peeters}). The above eigenvalue problem
is a matrix equation which, following a procedure analogous to
earlier works \cite{Berman2}, \cite{Trushin} and \cite{Peeters}, can be decoupled to a single ODE
of the second kind.

\section{\label{sec:level1} Separation of variables for the excitonic Hamiltonian}

Let us consider the excitons with zero center-of-mass momentum
$\pmb{K}=\pmb{k}_e+\pmb{k}_h=0$. Introducing $\pmb{k}\equiv
\pmb{k}_e = -\pmb{k}_h$, and using transformation to dimensionless polar coordinates 
$$
\p_x = \cos\varphi\; \p_r - \frac{\sin\varphi}{r}\,\p_\varphi, \quad
\p_y = \sin\varphi\; \p_r + \frac{\cos\varphi}{r}\,\p_\varphi,
$$
the exciton Hamiltonian (see Ref.~\cite{Peeters}) can be
written in the following way:

\begin{equation}
\begin{pmatrix}
-V(r) &  -he^{i\tau_h \varphi}D^-_h & -he^{-i\tau_e \varphi}D^+_e & 0 \\
-he^{-i\tau_h\varphi}D^+_h & \Delta_h-V(r) & 0 & -he^{-i\tau_e\varphi}D^+_e \\
-he^{i\tau_e\varphi}D^-_e & 0 & -\Delta_e-V(r) & -he^{i\tau_h\varphi}D^-_h \\
0 & -he^{i\tau_e\varphi}D^-_e & -he^{-i\tau_h\varphi}D^+_h &
\Lambda-V(r)
\end{pmatrix}.
\label{H}
\end{equation}

We define here differential operators $D^\pm_{e,h} \equiv
i\tau_{e,h}\p_r \pm \p_\varphi/r$ and quantities $\Delta_{e,h}
\equiv \Delta -\lambda s_{e,h}\tau_{e,h}$ (the effective band-gap, see \cite{Peeters}) with $\lambda$  being the spin-orbit coupling strength, $\Lambda \equiv
\lambda(s_e \tau_e-s_h\tau_h)$, and $h=a/r_0=v_F\hbar/(r_0 t)$ is the dimensionless parameter which is presumed to be small to apply WKB analysis. We have also expressed energy quantities
($\Delta$, $\lambda$, $V(r)$ etc.) in the hopping parameter $t$ units
 and the spatial variables in the screening length $r_0$ ones.

Due to spin splitting of the valence band, there are effectively two band gaps and, as a consequence, two different types of excitons. 
These are commonly referred to as A and B excitons. The method presented in this paper can be applied with both values of $s_h\tau_h=\pm 1$. When $s_h\tau_h=1$ the hamiltonian (\ref{H}) describes the A exciton, while $s_h\tau_h=-1$ relates to the B exciton. 
The A excitons in the $K$ and $K'$ valley have $s_e = 1$, $\tau_e =1$, 
$s_h=-1$, $\tau_h=-1$ and $s_e = -1$, $\tau_e =-1$, $s_h=1$, $\tau_h=1$, respectively. 
The B exciton has $s_e = -1$, $\tau_e =1$, $s_h=1$, $\tau_h=-1$. Note that the intervalley exciton, which can arise due to excitation of charge carriers with linearly polarized light, also has $s_e\tau_e = s_h\tau_h$. Under such conditions we obtain $\Lambda=0$. 

For the $A$ exciton in the $K$ and $K'$  valley (see \cite{Peeters}) we get
$$
\Delta_h=\Delta_e=\Delta-\lambda=\Delta_{s,\tau}.
$$
For the $B$ exciton in the $K$ valley (see \cite{Peeters}) we get
$$
\Delta_h=\Delta_e=\Delta+\lambda.
$$
Therefore, we can express $\Delta_h$ and $\Delta_e$ in terms of $\Delta_{s,\tau}$. In order to separate variables for the $A$ and $B$ excitons in the $K$ valley, we use the ansatz 
$$\psi(r,\varphi) = e^{i l \varphi} \left(
\psi_l^{(1)}(r) e^{-i\varphi}, \,\psi_l^{(2)}(r), \,\psi_l^{(3)}(r),
\,\psi_l^{(4)}(r) e^{i\varphi}\right)^T ,
$$
where $l\in{\bf Z}$ is an orbital quantum number. Thus, the
Hamiltonian~\eqref{H} acting on 
$$\psi_l(r) =\left(\psi_l^{(1)}(r),
\psi_l^{(2)}(r), \psi_l^{(3)}(r), \psi_l^{(4)}(r)\right)^T
$$ 
takes the form
\setlength\arraycolsep{7pt}
\renewcommand*{\arraystretch}{2}
\begin{equation}
\hat{H}_l =\begin{pmatrix}
-V(r) &  -hD_{-l} & hD_{-l} & 0 \\
-hD_l^\dagger & \Delta_{s,\tau}-V(r) & 0 & hD_{-l}^\dagger \\
hD_l^\dagger  & 0 & -\Delta_{s,\tau}-V(r) & -hD_{-l}^\dagger\\
0 & hD_l & -hD_l & -V(r)
\end{pmatrix},
\label{Hl}
\end{equation}
where $D_l=-i\p_r+\frac{il}{r}$, $D_l^\dagger=(-i\p_r)^\dagger+\frac{il}{r}$, $(-i\p_r)^\dagger = -i\p_r-i/r$, or, equivalently,
$\p_r^\dagger = -\p_r-1/r$. Note that the Hamiltonian  (\ref{Hl}) is
self-adjoint with respect to the scalar product
$$
\langle \pmb\psi, \pmb\chi \rangle = \sum_{i=1}^4 \int_0^\infty \psi^*_i(r) \chi_i(r)\, r dr.
$$

The spectral
problem
 \be
 \hat{H}_l\psi_{nl}(r)=E_{nl}\psi_{nl}(r),~~~n=0,1,2,...,
 \ee
for the Hamiltonian (\ref{Hl}) with the radial quantum number $n$
can be represented as the following system of four first-order ODE's
\be
 \begin{cases}
V_E\psi_l^{(1)}+hi\left(\p_r+\frac{l}{r}\right)\psi_l^{(2)}- hi\left(\p_r+\frac{l}{r}\right)\psi_l^{(3)}=0, \\
\left (\p_r-\frac{l-1}{r}\right)\psi_l^{(1)}-ih^{-1}(\Delta_{s,\tau}+V_E)\psi_l^{(2)}-
\\\left (\p_r+\frac{l}{r}\right)\psi_l^{(4)}=0,\\
\left (\p_r-\frac{l-1}{r}\right)\psi_l^{(1)}+ih^{-1}(-\Delta_{s,\tau}+V_E)\psi_l^{(3)}-
\\ \left (\p_r+\frac{l}{r}\right)\psi_l^{(4)}=0, \\
-ih\left (\p_r-\frac{l-1}{r}\right)\psi_l^{(2)}+
ih\left (\p_r-\frac{l-1}{r}\right)\psi_l^{(3)}+V_E\psi_l^{(4)}=0,
 \end{cases}
 \label{SODE1}
 \ee
where $V_E=-V(r)-E$. Transforming vector components
$\psi_l(r)$ into $\varphi_l^{(1,4)}=\psi_l^{(1)}\pm\psi_l^{(4)}$, $\varphi_l^{(2,3)}=\psi_l^{(2)}\pm\psi_l^{(3)}$,
we obtain a new system of ODE for the components of a new vector
$\varphi_l(r)=(\varphi_l^{(1)},\varphi_l^{(2)},\varphi_l^{(3)},\varphi_l^{(4)})^T$. This system is reduced into  a scalar second-order ODE

 \be
 \left(\varphi_l^{(3)}\right)''+\left(\frac{1}{r}-\frac{V'}{V_E}\right)\left(\varphi_l^{(3)}\right)'+\left(\frac{V_E^2-\Delta_{s,\tau}^2}{4h^2}-
 \frac{l^2}{r^2}\right)\varphi_l^{(3)}=0.
 \label{ODE1}
 \ee

To separate variables for the $A$ excitons in the $K'$ value we have to use the ansatz 
$$\psi(r,\varphi) = e^{i l \varphi} \left(
\psi_l^{(1)}(r) e^{i\varphi}, \,\psi_l^{(2)}(r), \,\psi_l^{(3)}(r),
\,\psi_l^{(4)}(r) e^{-i\varphi}\right)^T.
$$ 
As a result we obtain the same ODE (\ref{ODE1}).

We are going to calculate discrete spectrum of equation (\ref{ODE1}), assuming $E$ the spectral parameter. We also assume that $E>0$ as in this case the
quantity $E+V(r)$ does not vanish. Otherwise, we have to deal with singularities. This case requires further studies. The coefficients of (\ref{ODE1}) depend on
the spectral parameter $E$ nonlinearly. Eigenfunctions
$\varphi_l^{(3)}(r,E_n)$ are supposed to be square-integrable
($\varphi_l^{(3)}(r,E_n)\in L_2(0,+\infty)$). Taking into account that

$$V(r)=\frac{V_0}{r}+O \left(r^{-2} \right)$$
as $r\to +\infty$, the coefficients in
(\ref{ODE1}) could be simplified up to the leading order terms. This approximation results in the well-known Bessel-type ODE

$$
\left(\varphi_l^{(3)}\right)''+\frac{1}{r}\left(\varphi_l^{(3)}\right)'+ \left( \frac{E^2-\Delta_{s,\tau}^2}{4h^2}-
 \frac{l^2}{r^2} \right) \varphi_l^{(3)}=0.
 \label{ODEK}
$$

 As we seek a solution
decaying as $r\to +\infty$, the following inequality must hold
 \be
 0<E<\Delta_{s,\tau}.
 \label{ineqsp}
 \ee
As a result, we get the solution
 \be
 \varphi_l^{(3)}\sim
 K_l(k_Er),~~~~~k_E=\frac{\sqrt{\Delta_{s,\tau}^2-E^2}}{2h},
 \ee
where $K_l(r)$ is the MacDonald function that behaves as
$e^{-r}\sqrt{\frac{\pi}{2r}}$ when $r\to +\infty$. Thus, the energy levels must satisfy the inequality (\ref{ineqsp}).

Using the gauge transformation
 \be
 \varphi_l^{(3)}=w(r)\exp\left(-\frac{1}{2}\int \left(\frac{1}{r}-\frac{V'}{V_E} \right)dr\right)=w(r)\sqrt{\frac{V_E}{r}},
 \ee
we change the equation (\ref{ODE1}) into the following self-adjoint, canonical form
 \be
 w''+Q(r,E)w=0,
 \label{ODE2}
 \ee
 where
 $$
 Q(r,E)=\frac{V_E^2-\Delta_{s,\tau}^2}{4h^2}-
 \frac{l^2}{r^2}-\frac{1}{4}\left(\frac{1}{r}+\frac{V'}{V_E}\right)^2 -
 $$
 $$
 -\frac{1}{2}\left(-\frac{1}{r^2}+\frac{V''}{V_E}+\left(\frac{V'}{V_E}\right)^2\right).
 $$

 It is clear that, when $r\to +\infty$, we have $Q\sim -k_E^2$, and one of the two linearly independent solutions of (\ref{ODE2}) decays exponentially.

 Let us analyze the asymptotic behaviour of $w$ as $r\to 0$. For the case
 $l\neq 0$, up to the leading order, we may use the approximation
  \be
  Q(r,E)\sim \frac{1}{4r^2}-\frac{l^2}{r^2}.
  \ee
As a result we obtain
  \be
  w=C_1r^{1/2+l}+C_2r^{1/2-l}, ~~~~~C_{1,2}=const.
  \ee
Assuming $\varphi_l^{(3)}(r)$ bounded as $r\to 0$, we get $C_2=0$. Thus, for $l\ne 0$, we obtain that $w(0)=0$.

For the case
 $l=0$, up to the leading order, the following approximation may be used
  \be
  Q(r,E)\sim \frac{1}{4r^2}-\frac{3}{4r^2\log{r}^2}.
  \ee
Then the solution would behave as
 \be
 w\sim C\frac{r^{1/2}}{\sqrt{|\log{r}|}},~~~~~C=const.
 \ee
Thus, for $l=0$, we obtain again that $w(0)=0$.

As soon as compute $w(r)$ numerically, we can evaluate
$\varphi_l^{(3)}(r)$ and the other components of $\varphi^l(r)$
and then all components of $\psi^l(r)$. Namely, we have
 \be
 \varphi_l^{(1)}=\frac{2ilh}{r(E+V)}\varphi_l^{(3)},~~~~~\varphi_l^{(4)}=\frac{2ih}{r(E+V)}\left(\varphi_l^{(3)}\right)',
 \ee
 \be
 \varphi_l^{(2)}=\frac{\Delta_{s,\tau}}{E+V}\varphi_l^{(3)},
 \ee
and
 \be
 \psi_l^{(2,3)}=\frac{1}{2}\varphi_l^{(3)}\left(\frac{\Delta_{s,\tau}}{E+V}\pm 1\right).
 \ee

It is worth remarking that, if $l=0$,
we have
 \be
 \psi_l^{(1)}=-\psi_l^{(4)}=\varphi_l^{(4)}=\frac{2ih}{r(E+V)}\left(\varphi_l^{(3)}\right)',
 \ee
and if $l\ne 0$, then
 \be
 \psi_l^{(1)}=\frac{ih}{V_E}\left(\frac{l}{r}\varphi_l^{(3)}+\left(\varphi_l^{(3)}\right)'\right),
  \ee
  \be
  \psi_l^{(4)}=\frac{ih}{V_E}\left(\frac{l}{r}\varphi_l^{(3)}-\left(\varphi_l^{(3)}\right)'\right).
 \ee
 
\section{Semiclassical approach for $A$ and $B$ excitons}

In this section we solve the spectral problem for ODE (\ref{ODE1}) or
(\ref{ODE2}) for $r\in[0,+\infty]$ analytically using
semiclassical approximation (1D WKB approach) based on the method of
comparison equation that is well-known in the theory of
asymptotic methods in ODE (see for example \cite{Olver}, \cite{Fedoryk}, \cite{Borov}). It is worth remarking that semiclassical analysis has been successfully applied to many theoretical problems in graphene. We could mention briefly a few references: semiclassical approach to Berry phase analysis  \cite{Carmier}, the localized states \cite{Zalip1} and Dirac electron tunnelling in graphene \cite{Zalip2}, \cite{Zalip3}. When applying a semiclassical analysis, it is assumed that $h$ is a
small parameter ($h<<1$). For convenience, we also introduce a large
parameter $k=1/h>>1$.
 Let us consider first the case with $l\neq 0$. Let us rewrite 
 (\ref{ODE2}) in following way
  \be
  w''+k^2q(r,E)w=0,
  \label{ODE3}
  \ee
where
 $$
 q(r,E)=\frac{V_E^2-\Delta_{s,\tau}^2}{4}-h^2q_h(r,E),
 $$
 $$
 q_h=\frac{l^2}{r^2}+\frac{1}{4}\left(\frac{1}{r}+\frac{V'}{V_E}\right)^2+\frac{1}{2}\left(-\frac{1}{r^2}+\frac{V''}{V_E}+\left(\frac{V'}{V_E}\right)^2\right).
 $$
For this ODE we have got two turning points $r_{1,2}\in[0,+\infty]$,
$r_1<r_2$, for which $q(r_{1,2},E)=0$ that could locate close to each other. Thus, it is necessary  to employ uniform
asymptotic approximation relevant to the case when two turning points could coalesce (see \cite{Olver}, \cite{Fedoryk}).
The approximation involves the parabolic special function $D_{\nu}(x)$ (see \citep{Borov}, \cite{Abram}) that satisfies the
following ODE
 \be
 y''+\left(\nu+\frac{1}{2}-\frac{x^2}{4}\right)y=0,
 \ee
 and its asymptotic expansion for large $x$ is given by
 $$
D_{\nu}(x)\sim e^{-x^2/4}x^{\nu},~~~x\to +\infty.
 $$
Hence, the semiclassical approximation to the solution reads
 \be
 w_n(r)=\left(
 \frac{d-z^2}{q(r,E_n)}\right)^{\frac{1}{4}}D_{kd/2-1/2}(\sqrt{2k}z),
 \ee
 $$
 d=\frac{2}{\pi}\int\limits_{r_1}^{r_2}\sqrt{q(r,E_n)}dr.
 $$
The parameter $z$ is to be found from the following equations: for $r>r_2$ ($q(r,E_n)<0$, $z>\sqrt{d}$)
 $$
 \int\limits_{r_2}^r\sqrt{-q(r,E_n)}dr=
 $$
 \be
 \frac{1}{2}\left(z\sqrt{z^2-d}-d\log{(z+\sqrt{z^2-d})}\right)+\frac{d}{4}\log{d}.
 \ee
For $r_1<r<r_2$ ($q(r,E_n)>0$, $-\sqrt{d}<z<\sqrt{d}$), we have
 $$
 \int\limits_{r_1}^r\sqrt{q(r,E_n)}dr=
 $$
 \be
 \frac{1}{2}\left(z\sqrt{d-z^2}+d\arcsin{\frac{z}{\sqrt{d}}}\right)+\frac{\pi d}{4}.
 \ee
Finally, for $r<r_1$ ($q(r,E_n)<0$, $z<-\sqrt{d}$), $z$ is to be found from
 $$
 \int\limits_r^{r_1}\sqrt{-q(r,E_n)}dr=
 $$
 \be
 \frac{1}{2}\left(-z\sqrt{z^2-d}-d\log{(-z+\sqrt{z^2-d})}\right)+\frac{d}{4}\log{d}.
 \ee
The discrete spectrum $\{E_n\}_{n=0}^N$ is obtained from Bohr-Sommerfeld quantization rule (see \cite{Olver}, \cite{Fedoryk})
 \be
 \int\limits_{r_1}^{r_2}\sqrt{q(r,E_n)}dr=\pi h\left(n+\frac{1}{2}\right).
 \label{QC1}
 \ee

However, the discussed above form of WKB approximation is not
applicable to describe the case $l=0$. In this situation for
(\ref{ODE3}), we have got only one turning point $r_1$, and 
the integral in the corresponding Bohr-Sommerfeld quantization rule diverges due to the
singularity of $q(r,E)\sim 1/(4r^2)$ as $r\to 0$.

Thus, for the case $l=0$, WKB analysis is to be developed in a
different way. For ODE (\ref{ODE1})
the change of variable $r=e^t$ ($t\in (-\infty,+\infty)$) results
in
 \be
 (\varphi_l^{(3)})''-\frac{d}{dt}\log{(-V_E)}(\varphi_l^{(3)})'+e^{2t}\frac{(E+V)^2-\Delta_{s,\tau}^2}{4h^2}\varphi_l^{(3)}=0.
 \label{ODE5}
 \ee

Introducing a new dependent variable
 \be
 u(t)=\frac{\varphi_l^{(3)}(r)}{\sqrt{E+V(r)}},
 \ee
we obtain
 \be
 u''+k^2p(t,E)u=0,~~~ k>>1,
 \label{ODE6}
 \ee
where
 \be
p(t,E)=e^{2t}\frac{(E+V)^2-\Delta_{s,\tau}^2}{4}-h^2\left(\frac{V''}{2V_E}+\frac{3V'^2}{4V_E^2}
\right).
 \ee
In this expression all derivatives are assumed to be evaluated with respect to
the variable $t$.

For this ODE (\ref{ODE6}), we have got two turning points
$t_{1,2}\in[-\infty,+\infty]$, $t_1<t_2$, for which $p(t_{1,2},E)=0$
that are distinct from each other. Thus, it is reasonable to
employ uniform asymptotic approximation for the case when two turning points
do not coalesce (see \cite{Olver}, \cite{Fedoryk}), with the Airy special function
$v(x)=\sqrt{\pi}Ai(x)$ that satisfies the following ODE $y''-xy=0$ (\citep{Borov}, \citep{Abram}).
Hence, we obtain for $t<t^*$
 \be
 u_n(t)=C_0k^{1/6}\left(\frac{z_1(t)}{p(t,E_n)}\right)^{1/4}v\left(-k^{2/3}z_1(t)\right),
 \label{U1}
 \ee
 and for $t>t^*$
 \be
 u_n(t)=C_0(-1)^nk^{1/6}\left(\frac{z_2(t)}{p(t,E_n)}\right)^{1/4}v\left(-k^{2/3}z_2(t)\right),
 \label{U2}
 \ee
where $t^*$ is an arbitrary point between $t_1$ and $t_2$, and
$C_0=const$. For these asymptotic formulae, the real valued functions $z_{1,2}(t)$ are defined by
 \be
 z_1(t)=-\left(\frac{3}{2}\int\limits_t^{t_1}\sqrt{-p(t,E_n)}dt
 \right)^{2/3}<0,~~~~~t<t_1,
 \ee
 \be
 z_1(t)=\left(\frac{3}{2}\int\limits_{t_{1}}^t\sqrt{p(t,E_n)}dt
 \right)^{2/3}>0,~~~~~t_1<t<t^*,
 \ee
 \be
 z_2(t)=\left(\frac{3}{2}\int\limits_t^{t_2}\sqrt{p(t,E_n)}dt
 \right)^{2/3}>0,~~~~~t^*<t<t_2,
 \ee
 \be
 z_2(t)=-\left(\frac{3}{2}\int\limits_{t_{2}}^t\sqrt{-p(t,E_n)}dt
 \right)^{2/3}<0,~~~~~t>t_2.
 \ee
It is worth noting that  $p(t,E_n)<0$ for $t<t_1$,
$p(t,E_n)>0$ for $t_1<t<t_2$, and $p(t,E_n)<0$ for $t>t_2$.

The descrete spectrum $\{E_n\}_{n=0}^N$ in this case is again
obtained from the Bohr-Sommerfeld quantization rule
 \be
 \int\limits_{t_1}^{t_2}\sqrt{p(t,E_n)}dt=\pi h\left(n+\frac{1}{2}\right).
 \label{QC0}
 \ee
This quantization rule makes both asymptotic formulae
(\ref{U1}) and (\ref{U2}) consistent.

\section{$A$ exciton in $K$ valley - numerical analysis}

We confine our numerical analysis by considering the special case $s_e=\tau_e=1$, $s_h=\tau_h=-1$ which
corresponds to $A$ exciton in $K$ valley (see \cite{Peeters}) and was studied theoretically in the previous sections. 
On the basis of ODE (\ref{ODE2}), by truncating the semi-axes to the
segment $[0,R]$, we solve our spectral problem using finite differences method (FDM) with following boundary
conditions
 \be
 w(0)=w(R)=0.
 \ee

We take into account that the eigenfunction must decay exponentially. Thus, the discrete version of (\ref{ODE2}) is the homogeneous
system of linear algebraic equations that involves the energy spectral parameter in a nonlinear way
 \be
 A(E)\bar w=0.
 \label{FDSys}
 \ee
The system is formed by the  tridiagonal matrix $m\times m$ which entries are given by
$$
A_{i,i}(E)=4-2Q_ih_r^2,~~~A_{i,i+1}(E)=-h_r-2,
$$
$$
A_{i+1,i}(E)=h_r-2,
$$
where $r_i=ih_r$, $i=1,2,...,m$, $h_r=R/(m+1)$, $Q_i=Q(r_i)$, $\bar
w=(w_1,w_2,...,w_m)^T$, $w_i=w(r_i)$. The discrete excitonic
spectrum is determined approximately by the equation $\det(A(E))=0$,
which could be solved by iterative Newton method.

The described above numerical method is relatively simple and straightforward.
Various tests demonstrated its stability with the increase of the size of the system $m$ and the parameter $R$. 
 
\begin{figure}
\includegraphics[width=1.07\linewidth]{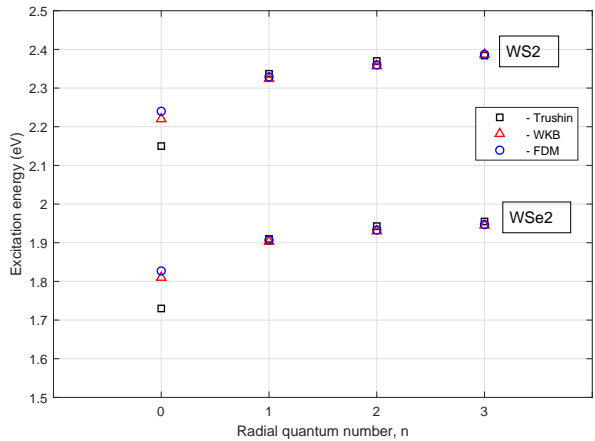}
\caption{The excitonic spectrum $E_{n0}$ for the
s-states ($l=0$) with  $n=0,1,2,3$ for two examples of WS2 and WSe2. The data of Trushin analytical solution (black squares) are shown for comparison together
with  the WKB approximation values (red triangles) and the finite difference method data (FDM, blue circles).}\label{fig1}
\end{figure}

Fig. 1 shows comparison between the data for the excitonic spectrum $E_{n0}$
for the s-states ($l=0$) with  $n=0,1,2,3$ obtained in \cite{Trushin}, and
the values computed on the basis  of  WKB approximation  and FDM data. It has been done  first
for the example of WS2 with the parameters
$\Delta_{s,\tau} = 2.4 eV$, $a = 3.197A^0$, $t = 1.25 eV$,  and then, for the example of WSe2 with
 $\Delta_{s,\tau} = 1.97 eV$, $a = 3.317A^0$ , and $t = 1.13
eV$. Both examples involve the SiO2
substrate with permittivity $\epsilon=3.9$. For both
examples we have $h=0.1157$. It is
clear that the ground state energy obtained with FDM and WKB differs from analytical solution. The discrepancy of excited states, however, is relatively low . We have to take into account that semiclassical approximation works perfectly for relatively large quantization indices. It is worth remarking that analytical solution, as it was demonstrated in \cite{Trushin} agrees with the experimental data obtained in \cite{Chernik} and \cite{He}.

\begin{figure}
\includegraphics[width=1.05\linewidth]{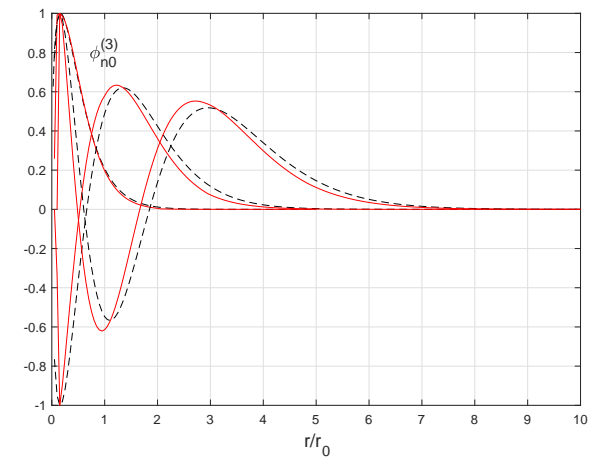}
\caption{FDM data (black dashed curves) versus WKB
approximation values (red continuous curves) for the case $n=0,1,2$ and $l=0$ for WS2 example showing
 $\varphi_{n0}^{(3)}$ dependence on $r$.}\label{fig2}
\end{figure}

\begin{figure}
\includegraphics[width=1.05\linewidth]{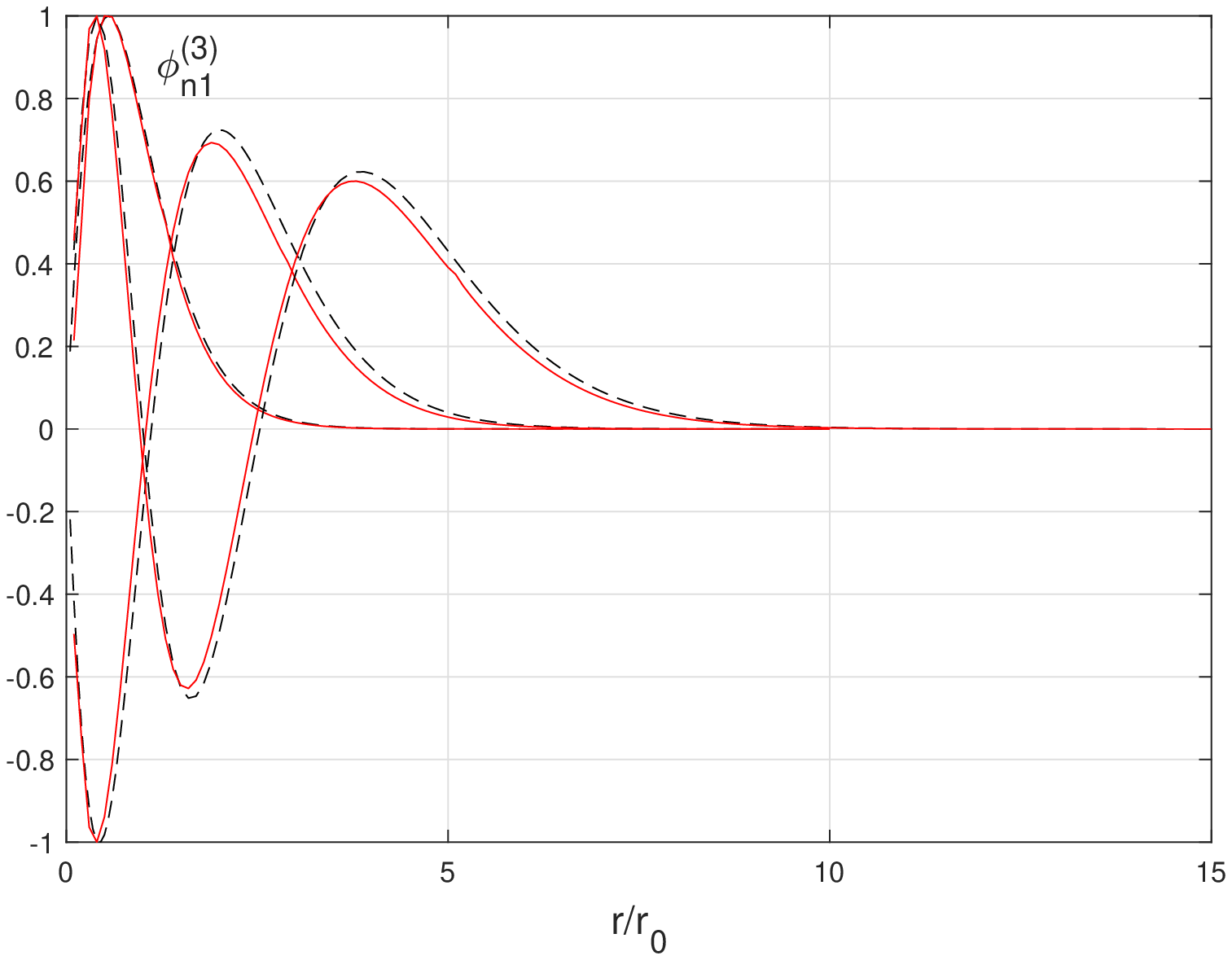}
\caption{FDM data (black dashed curves) versus WKB
approximation values (red continuous curves) for the case $n=0,1,2$ and $l=1$ for WS2 example showing
 $\varphi_{n1}^{(3)}$ dependence on $r$.}\label{fig3}
\end{figure}

\begin{figure}
\includegraphics[width=1.05\linewidth]{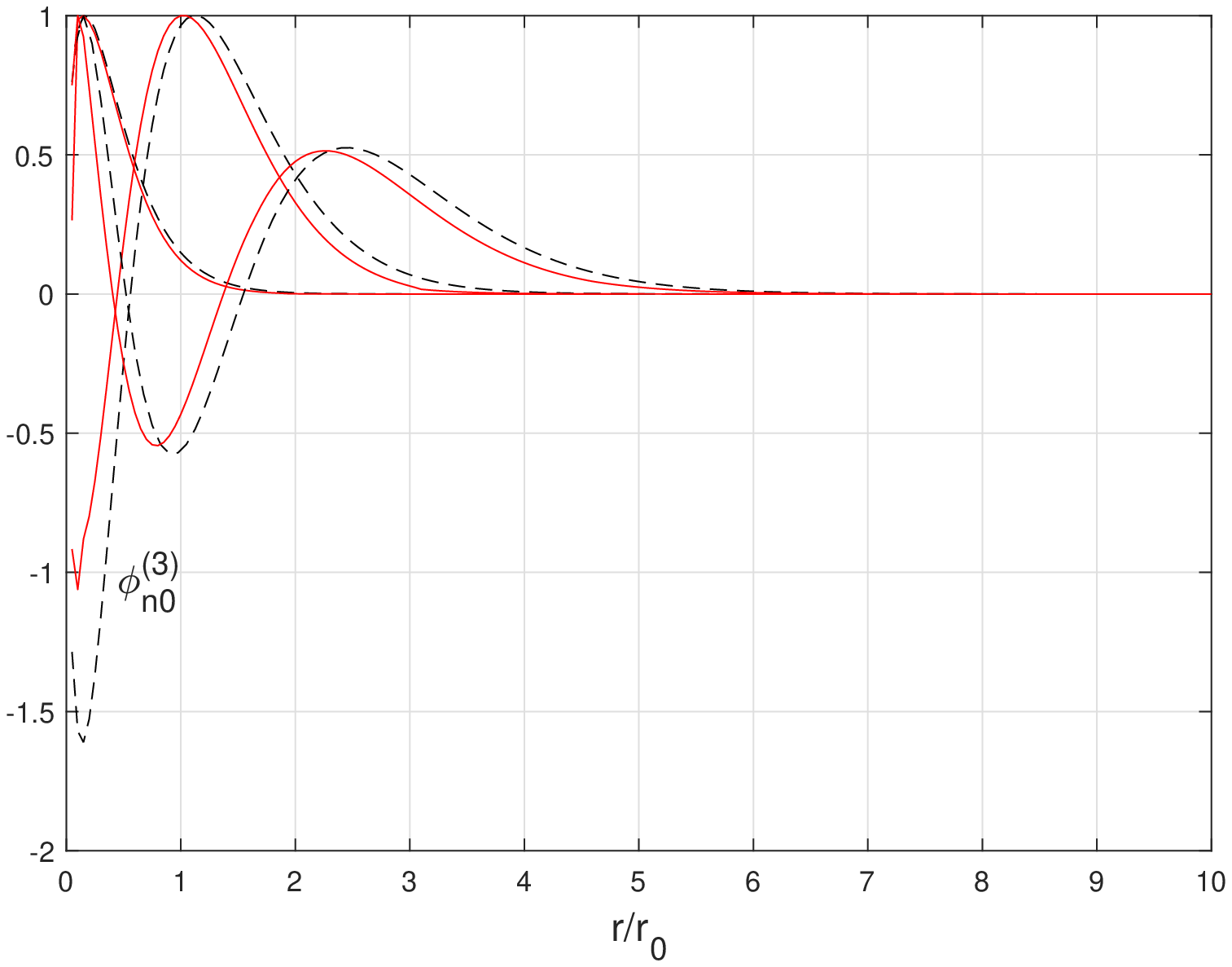}
\caption{FDM data (black dashed curves) versus WKB
approximation values (red continuous curves) for the case $n=0,1,2$ and $l=0$ for WSe2 example showing
 $\varphi_{n0}^{(3)}$ dependence on $r$.}\label{fig4}
\end{figure}

\begin{figure}
\includegraphics[width=1.05\linewidth]{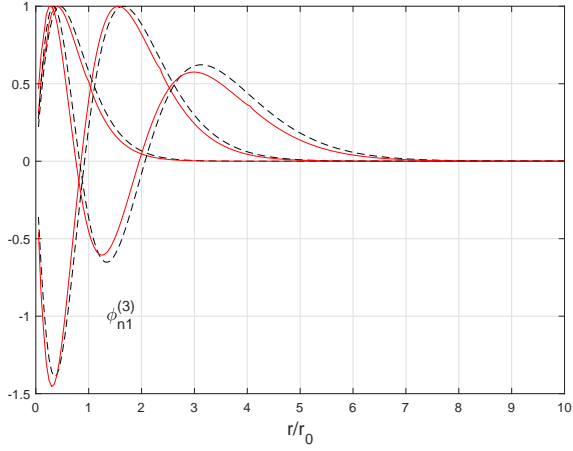}
\caption{FDM data (black dashed curves) versus WKB
approximation values (red continuous curves) for the case $n=0,1,2$ and $l=1$ for WSe2 example showing
 $\varphi_{n1}^{(3)}$ dependence on $r$.}\label{fig5}
\end{figure}

Below we compare the data obtained by means of WKB approximation and FDM  for two examples of WS2 and WSe2 with the parameters described above. Thus in Fig. 2 and Fig. 3 for 
$l=0, 1$, correspondingly, the behaviour of
$\varphi_{nl}^{(3)}(r)$ with $n=0,1,2$ is shown. The  WKB
approximation values are represented with red continuous curves, whereas FDM data are shown with black dashed curves. The values of $E_{n0}$  for the s-states are shown in Fig 1. For the p-states, using WKB approximation, we obtain $E_{01}=2.304eV$,  $E_{11}=2.351eV$, $E_{21}=2.371eV$. The corresponding values computed with the help of FDM are $E_{01}=2.310eV$, $E_{11}=2.353eV$, $E_{21}=2.371eV$.  It is worth remarking that, for the values of $n=1,2$, the curves intersect the $r$ axis one and two times correspondingly, opposite to the case with $n=0$ without intersection. In this case one could clearly observe the compliance of data. 

Similarly, in Fig. 4 and Fig. 5, again for 
$l=0, 1$ correspondingly,  the behaviour of
$\varphi_{nl}^{(3)}(r)$ with $n=0,1,2$ also demonstrates good agreement between the WKB and FDM data. The values of $E_{n0}$  for the s-states are shown in Fig 1. For the p-states, using WKB approximation, we obtained $E_{01}=1.877eV$,  $E_{11}=1.921eV$, $E_{21}=1.939eV$. The corresponding values computed with the help of FDM are $E_{01}=1.882eV$, $E_{11}=1.923eV$, $E_{21}=1.940eV$. Thus, we have illustrated  that both methods could provide reliable results.

\begin{figure}
\includegraphics[width=1.05\linewidth]{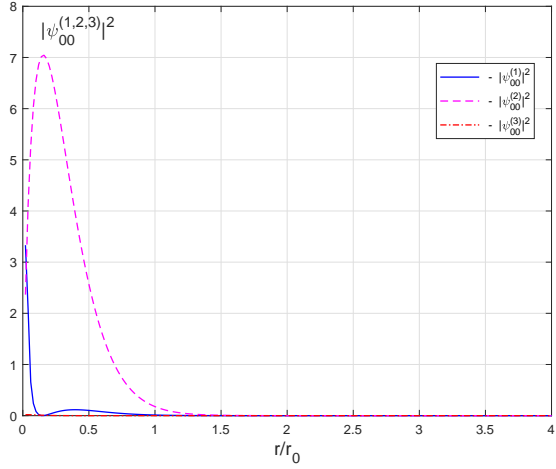}
\caption{Semiclassical approximation data for the probability distributions for the case $n=0$ and $l=0$ for WS2 example showing
 $|\psi_{00}^{(1,2,3)}|^2$ dependence on $r$.}\label{fig6}
\end{figure}

\begin{figure}
\includegraphics[width=1.05\linewidth]{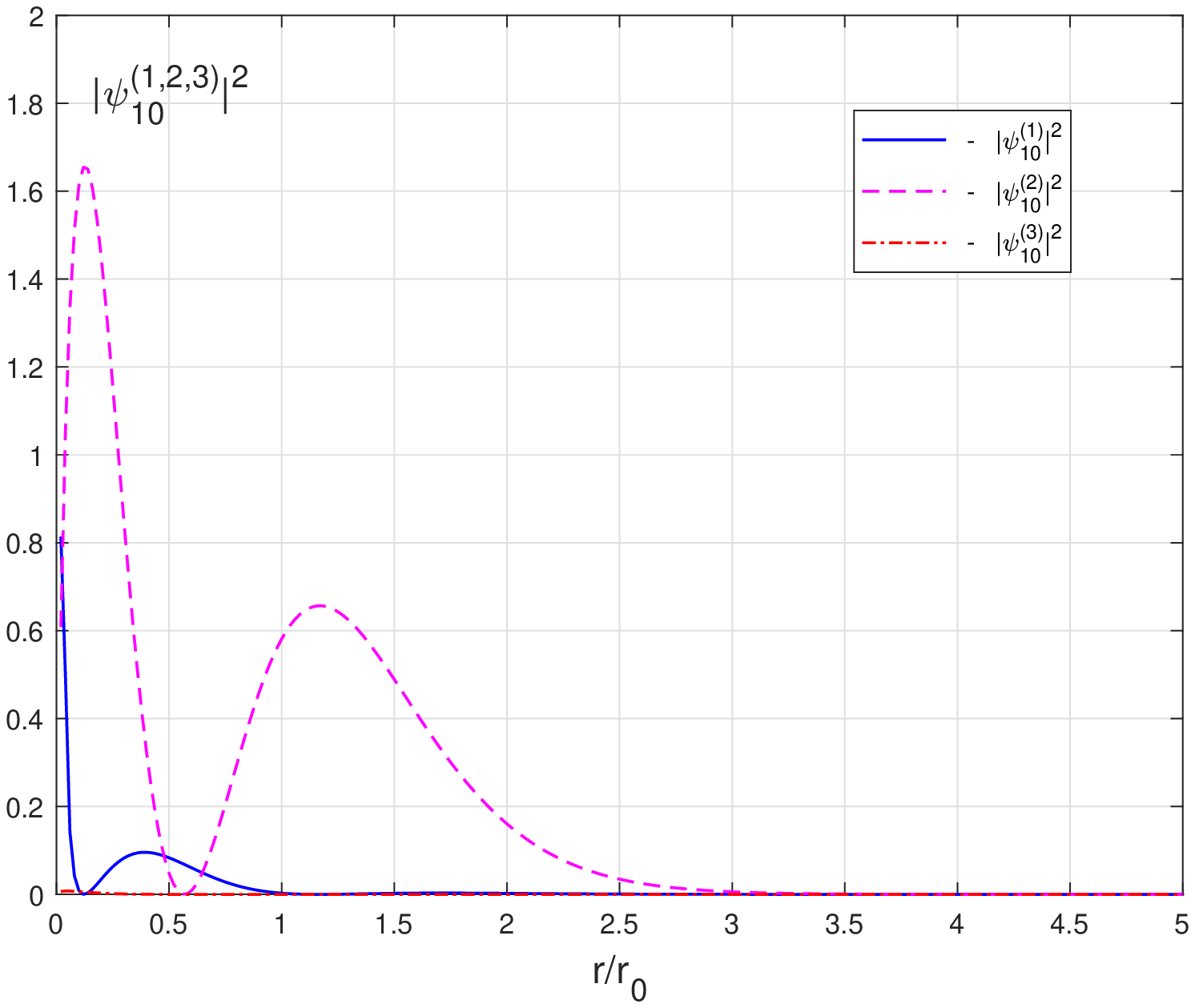}
\caption{Semiclassical approximation data for the probability distributions for the case $n=1$ and $l=0$ for WS2 example showing
 $|\psi_{10}^{(1,2,3)}|^2$ dependence on $r$.}\label{fig7}
\end{figure}

\begin{figure}
\includegraphics[width=1.05\linewidth]{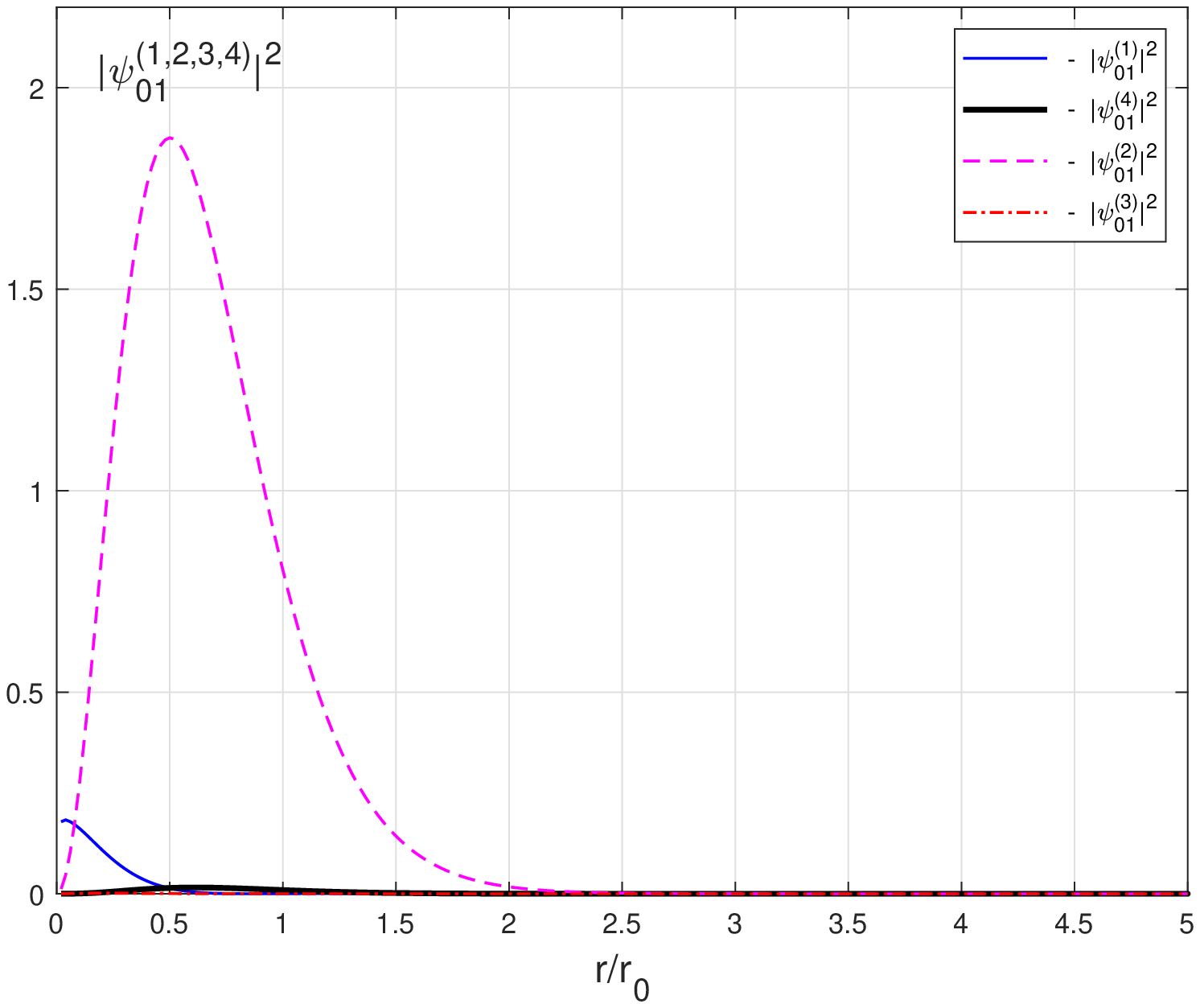}
\caption{Semiclassical approximation data for the probability distributions for the case $n=0$ and $l=1$ for WS2 example showing
 $|\psi_{01}^{(1,2,3,4)}|^2$ dependence on $r$.}\label{fig8}
\end{figure}

\begin{figure}
\includegraphics[width=1.05\linewidth]{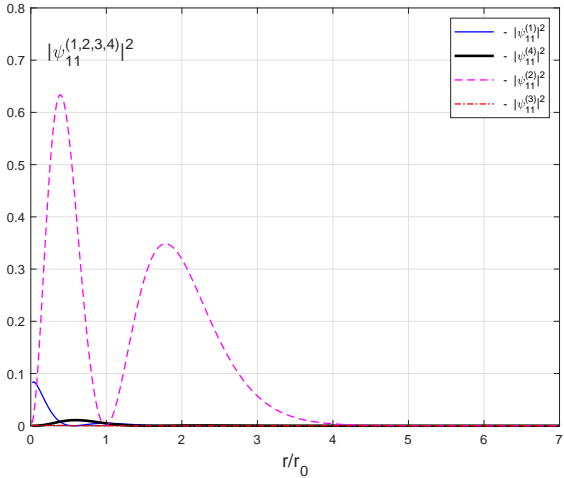}
\caption{Semiclassical approximation data for the probability distributions for the case $n=1$ and $l=1$ for WS2 example showing
 $|\psi_{11}^{(1,2,3,4)}|^2$ dependence on $r$.}\label{fig9}
\end{figure}

In Fig. 6-9  for the WS2 example  the dependence of $|\psi_{n0}^{(1,2,3)}|^2$ and $|\psi_{n1}^{(1,2,3,4)}|^2$ on $r$ for $n=0,1$ 
is demonstrated. All four figures show that the component 
$|\psi_{n0}^{(3)}|^2$ prevails with respect to the others.

\section{Conclusion}

The energy spectrum of excitons in monolayer
transition metal dichalcogenides was calculated using  a
multiband model. In this model, we used the excitonic
Hamiltonian in the product base of single-particle states at the
conduction and valence band edges constructed in \cite{Peeters}. Following
the separation of variables, we decoupled the corresponding system of the first order ODE for the radial eigen-vector components
and solved the resulting second order ODE
using the finite difference method. Thus, we determined the energy
levels of the electron-hole pairs and the corresponding
eigen-states. We also developed WKB approach to solve this spectral
problem in semiclassical approximation for the resulting ODE and demonstrated a very good agreement between the numerical data obtained by both methods. We also compared our  results for the energy spectrum  with
other theoretical works for excitons.

\begin{acknowledgments}
This work is supported by the Russian Science Foundation under Grant No. 18-12-00429.
The authors would like to thank Dr Alexei Vagov, Dr Dmitry R. Gulevich and Yaroslav V. Zhumagulov for helpful discussions.
%for the constructive discussions and valuable remarks.
%\dots.
\end{acknowledgments}

\bibliography{diracextn1bib}% Produces the bibliography via BibTeX.

\end{document}